\documentclass[a4paper]{spie}
\usepackage{cite}
\usepackage{amsmath,amssymb,amsfonts}
\usepackage{algorithmic}
\usepackage{graphicx, epstopdf}
\usepackage{subcaption}
\usepackage{textcomp}
\usepackage{booktabs}
\usepackage{siunitx}
\usepackage{enumitem}

\title{Automatic Plane Adjustment of Orthopedic Intra-operative Flat Panel Detector CT-Volumes}
\author{Celia Mart\'{i}n Vicario\supit{a}, Florian Kordon\supit{a}\supit{b}\supit{c}, Felix Denzinger\supit{a}\supit{c}, Jan Siad El Barbari\supit{d}, Maxim Privalov\supit{d}, Jochen Franke\supit{d}, Sarina Thomas\supit{e}, Lisa Kausch\supit{e}, Andreas Maier\supit{a}, and Holger Kunze\supit{c}\supit{a}
\skiplinehalf
\supit{a}Pattern Recognition Lab, Friedrich-Alexander-Universität Erlangen-Nürnberg, Erlangen, Germany\\
\supit{b}Erlangen Graduate School in Advanced Optical Technologies, Friedrich-Alexander-Universität Erlangen-Nürnberg (FAU), Erlangen, Germany\\
\supit{c}Siemens Healthcare GmbH, Forchheim, Germany\\
\supit{d}Department for Trauma and Orthopaedic Surgery, BG Trauma Center Ludwigshafen, Ludwigshafen\\
\supit{e}Division of Medical Image Computing, German Cancer Research Center, Heidelberg, Germany
}

\authorinfo{Further author information: (Send correspondence to C. Mart\'{i}n Vicario)\\C. Mart\'{i}n Vicario: E-mail: celia.martin@fau.de}

\begin{document}
\maketitle

\begin{abstract}
\noindent\textbf{Purpose}
3D acquisitions are often acquired to assess the result in orthopedic trauma surgery. With a mobile C-Arm system, these acquisitions can be performed intra-operatively. That reduces the number of required revision surgeries. However, due to the operation room setup, the acquisitions typically cannot be performed such that the acquired volumes are aligned to the anatomical regions. Thus, the multiplanar reconstructed (MPR) planes need to be adjusted manually during the review of the volume. In this paper, we present a detailed study of multi-task learning (MTL) regression networks to estimate the parameters of the MPR planes.

\noindent\textbf{Approach}
First, various mathematical descriptions for rotation, including Euler angle, quaternion, and matrix representation, are revised. Then, three different MTL network architectures based on the PoseNet are compared with a single task learning network.

\noindent\textbf{Results}
Using a matrix description rather than the Euler angle description, the accuracy of the regressed normals improves from $7.7^{\circ}$ to $7.3^{\circ}$ in the mean value for single anatomies. The multi-head approach improves the regression of the plane position from $7.4 \si{mm}$ to $6.1\si{mm}$, while the orientation does not benefit from this approach.

\noindent\textbf{Conclusions}
The results show that a multi-head approach can lead to slightly better results than the individual tasks networks. The most important benefit of the MTL approach is that it is a single network for standard plane regression for all body regions with a reduced number of stored parameters.

\end{abstract}


\section{Introduction}
\label{sec:introduction}
The default imaging modality to assess fracture reduction, implant position, and overall outcome during an orthopedic trauma surgery is X-ray imaging. However, the situation cannot be clearly judged from the X-ray image in complex anatomical regions like calcaneus, ankle, wrist, or knee in complex anatomical regions like calcaneus, ankle, wrist, or knee. This is due to ambiguities caused by overlapping or convex bones so that the positions of implants with respect to the corresponding bones are difficult to judge. Therefore, the acquisition of 3D scans is recommended before releasing the patient from the hospital. If 3D imaging is performed post-operatively, e.g., using a diagnostic CT system, not every minor finding will lead to revision surgery to spare the patient the risks of additional surgery. However, recent studies have shown that intra-operative 3D imaging has led to corrections of up to $40\,\%$ of surgeries, depending on the body region \cite{atesok_use_2007,beck_benefit_2009,beisemann_intraoperative_2019,carelsen_does_2008,franke_intraoperative_2012,franke_intraoperative_2014,gwak_intraoperative_2015,keil_intraoperative_2019,kendoff_intraoperative_2009,schnetzke_intraoperative_2018}. Thus intra-operative 3D imaging reduces the number of revision surgeries and improves the outcome of  surgeries as also minor findings are usually corrected. 

For intra-operative acquisition of 3D volumes, mobile C-arm systems are usually used, which are capable of cone-beam tomography (CBCT). These systems typically have a relatively limited field of view with a volume edge length of about $160\,\si{mm}$ to $250\,\si{mm}$. Consequently, the captured anatomy section and thus the anatomical landmarks' position and visibility may vary substantially.

When reading a 3D volume, the volume should be aligned to the anatomical structures in a standardized way as it is done in the radiology department. The key slices that contain anatomical structures which are decisive for the assessment of intervention results are called standard planes. Typically there are three of them: the axial, coronal, and sagittal plane. From an intra-operative 3D volume, they are typically obtained by the multiplanar reconstruction (MPR) technique. Generally, the three planes are orthogonal to each other, but in some regions, instead of these three orthogonal planes, an oblique plane provides the required information. One example of an oblique plane is the semi-coronal plane in the calcaneus region, a modification of the coronal plane that is not orthogonal to the axial and sagittal planes and which allows the evaluation of the reconstruction of the posterior talar surface \cite{grutzner_rontgenhelfer_2004}.

In Kausch et al.\cite{kausch_toward_2020} it has been shown that the accuracy of surgeons adjusting the standard MPRs highly depends on the region. In the lumbar spine region, where the planes can be adjusted using well-defined landmarks, the inter-rater difference was about half compared to the proximal femur region, where these kinds of landmarks are missing. The  mean inter-rater variance was measured up to $6.3^{\circ}$ for the normals and up to $9.3\,\si{mm}$ for the plane position.

As mobile C-Arms systems lack information about the spatial relationship between the system and the anatomical region, the adjustment of the plane position and orientation needs to be performed at the workstation in the operating room. This alignment of the planes is a manual task which takes $46$ to $210$ seconds depending on the experience level of the surgeon and thus, is a time-consuming step in a surgery \cite{brehler_intra-operative_2015,brehler_intra-operative_2016}.

Slice alignment in acquired volumes is a rather old topic. While the initial focus was on automatic rotation of the brain CT \cite{hirshberg_evaluating_2011,tan_approach_2019,qi_ideal_2013-1}, with the invention of 3D capable mobile C-arms systems -- which were used mainly in orthopedic and trauma surgery environments -- also other body parts like extremities attracted increased attention in research. Speeded Up Robust Features (SURF) were used in Brehler et al.\cite{brehler_intra-operative_2016} to register the acquired volume with an atlas that has annotated MPR planes. This method requires careful choice of the atlas and feature extraction method, but even then this approach has a limited capture range of rotation. Therefore, in Thomas\cite{thomas_thesis_2020} shape models with attached labels for the MPR planes were used. For generating the shape models, multiple volumes need to be manually segmented, which is time-consuming work. 

To account for small volume sizes that lead to cropped bones, and to be invariant to different metal implants positions, much effort and domain knowledge during the registration was applied to obtain a robust algorithm for one region. This leads to a long execution time of $23\,\si{s}$ for the shape model registration and the subsequent plane regression.

Artificial intelligence systems allow performing this task in a considerably faster time. An active research field for standard plane regression task is ultrasound imaging, for which in Lu et al.\cite{lu_autompr_2008} probabilistic boosting trees are used to estimate 9 transform parameters of the target MPRs using a multi-stage approach. Li et al.\cite{li_standard_2018} propose an iterative approach where a CNN repeatedly estimates the transform between a 2D plane and the standard plane. Using this approach, they can circumvent a fully 3D approach as only a small number of plane samples and updates are necessary until the regression converges.

In a more general sense, spatial transformer networks (STN) \cite{jaderberg_2015} predict the parameters of an affine transform matrix that is used to manipulate feature maps in a convolutional architecture spatially. No direct supervision for the transform is used, allowing the network to optimize towards a spatial configuration that maximizes the performance of the actual supervised target task.
The $\Omega$-Net by Vigneault et al.\cite{VIGNEAULT201895} modifies this approach by estimating the transform parameters for direct manipulation of the input image data. Conditional on the feature maps of a prior segmentation CNN, direct ground truth for the transformation parameters is used to bring the input images to a canonical form that better suits the downstream segmentation task.

Martin et al.\cite{martin_vicario_automatic_2020} uses a Pose-Net for the regression of the plane parameters. These plane parameters can be interpreted as transformation parameters. Comparing the structure of the Pose-Net with that of the STN, it can be clearly seen that the convolutional layers resemble the localization network, and the fully connected layers resemble the final regression layer. Thus, Martin et al.\cite{martin_vicario_automatic_2020} avoid the additional overhead of the segmentation introduced by the $\Omega$-Net while retaining the approach of supervising the transform parameters, which are of interest for the current task.

This article contributes in multiple ways:
\begin{itemize}[label={--}]
    \item We extend our initial ablation study presented in Martin et al.\cite{martin_vicario_automatic_2020} to give better insight in the performance of the algorithm. 
   
    \item We analyze different multi-task learning (MTL) approaches to improve the performance of the baseline algorithm. Typically, the number of available volumes per body region is small. Caruana et al.\cite{caruana_multitask_1997} showed that multi-task learning (MTL) can help to find the right shared representation for related tasks when only little data are available for the single tasks. MTL also helps to handle overfitting issues. Baxter \cite{baxter_bayesianinformation_1997} showed that parameter sharing reduces this risk substantially. Therefore, simultaneous learning for several tasks can help to find more appropriate representations and thus reduce the risk of overfitting. Furthermore, such combined training of MPR regression for different body regions can help to improve regression performance. We want to make use of this property of MTL in this work.
    
    The approach of MTL also has a practical benefit: single task networks are stored separately for each anatomical region and need to be loaded on demand. Measurements have shown that it takes up to $1\,\si{s}$ to load them from a hard drive to a graphics card. A combined network for which the parameters can be loaded once and then stay in memory would be beneficial. Therefore, we compare the results of region-specific networks to combined networks with and without knowledge about the body region and a multi-head approach. 
    
    \item We increase the number of evaluated body regions by adding proximal tibia (knee) and distal radius (wrist) to calcaneus and ankle, and we also show the results of an additional plane orientation representation and compare it to the already published results of Martin et al. \cite{martin_vicario_automatic_2020}.
\end{itemize}

In Section \ref{sec:methods} we present the employed mathematical description of planes, including a newly introduced second version of the 6D method \cite{zhou_continuity_2019}. We describe the normalization of the coordinate system and introduce the different neural network architectures we want to compare. Furthermore, the cost function for optimization is introduced. The implementation and the data we used for training and testing, as well as the study design, are described in Section \ref{sec:experiments}. After that, we present the results of our experiments and discuss the results in Section \ref{sec:results}.

\section{Methods}
\label{sec:methods}
\subsection{Plane Description}
An MPR plane can be described by its center position $A$ and the vectors $e_u$ and $e_v$ showing in the directions of the rows and columns. Its normal $e_w$ is the cross product of these two directions. 

For each plane, a rigid homogeneous transformation $T$ from the volume coordinate system to the plane coordinate system exists which can be decomposed into a 3x3 rotation matrix $R$ and a 3-element translation  $t$. 
\begin{equation}
  T = \left[ 
	\begin{matrix}
		R & t \\
		0 & 1
	\end{matrix}
	  \right]
\end{equation}
The 9 parameters of the $3 \times 3$ rotation matrix are highly coupled. So, the column vectors are normalized, the dot product of two vectors is zero, and one column vector can be calculated by the cross product of the other two vectors. These properties are utilized by the 6D method \cite{zhou_continuity_2019}. With this method, the values of two vectors are estimated by the neural network. Typically, the first two columns are utilized. However, it might also be favorable to regress the first and the third column instead of the second column as it encodes the normal of the plane which itself is part of the score function (Equ. \ref{equ_weight}), which will be introduced below. We denote the 6D method which regresses the parameters for x and y direction with $\textrm{6D}_{xy}$ and the one which regresses the x and z direction with $\textrm{6D}_{xz}$. After regression of the values, each column vector is normalized and the missing column vector is calculated as the cross product. As the matrix is a pure rotation matrix, its entries are in the range of $\left[-1,\,1\right]$.

A more common way to regress rotation parameters is to decompose the matrix into Euler angles or use a unit quaternion representation. While Euler angles suffer from discontinuous values, the quaternion representation does not have this problem. To overcome the limitation for Euler angles, we do not regress directly the angular value but their sine and cosine values. The actual angle value is then calculated from the regressed values using the $\textrm{atan2}$ method. Another advantage of this method is that the parameter range of the values is compressed to the range $\left[-1,\,1\right]$. The same range applies to the values of the quaternions.

The translation is normalized with respect to the volumes' dimensions and thus also lies in the range of $\left[-1,\,1\right]$ with the origin placed at the center of the volume.

\subsection{Separate and Combined Networks}
In Martin et al.\cite{martin_vicario_automatic_2020} separate networks were used for different anatomical regions. For each region, a single network for the regression of all three plane parameters achieved the best performance. However, it was not analyzed how one single network for all body regions performs. 

\begin{figure}[t]
\centering
\caption{Schematic visualization of the analyzed network architectures. (\subref{fig:Base_Network}) Baseline network without providing body region information to the network. The five convolutional blocks consist of a 3D convolutional layer (red), followed by a ReLU activation function (orange), batch normalization (brown), and a max pooling operation (yellow). The obtained features are fed into three fully connected layers (green). (\subref{fig:Class_Network}) Combined network architecture for all body regions with additional information about the body region fed into the first fully connected layer. (\subref{fig:Multihead_Network}) Multi-head network, convolutional blocks are shared across body regions, individual fully connected layers for the different body regions.}
\begin{subfigure}[l]{0.19\linewidth}
\centering
\includegraphics[height=3.4cm]{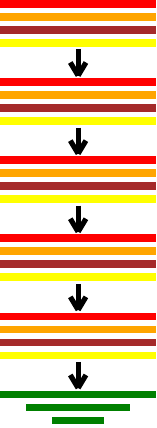}
\caption{}
\label{fig:Base_Network}
\end{subfigure}
\begin{subfigure}[l]{0.19\linewidth}
\centering
\includegraphics[height=3.4cm]{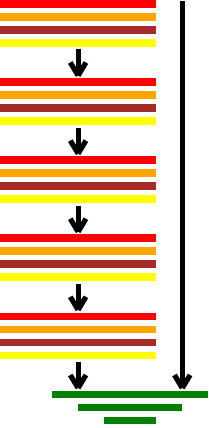}
\caption{}
\label{fig:Class_Network}
\end{subfigure}
\begin{subfigure}[l]{0.60\linewidth}
\centering
\includegraphics[height=3.4cm]{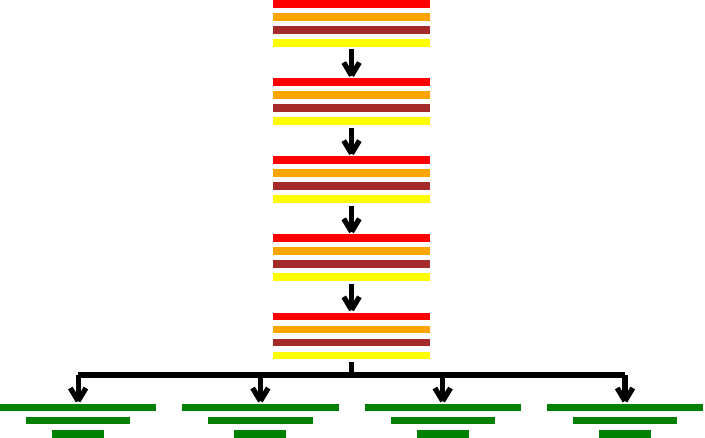}
\caption{}
\label{fig:Multihead_Network}
\end{subfigure}

\end{figure}

In preliminary experiments we compared the performance of VGG-16\cite{Simonyan2015}, ResNet-34\cite{He2016}, and PoseNet\cite{Bui_PoseNet} alike networks. We could observe the PoseNet to generalize better and to be more robust compared to the other two architectures. Therefore, we chose the PoseNet  as baseline network for our study (Figure \ref{fig:Base_Network}).

The PoseNet network consists of 5 convolutional layers and 3 fully connected layers. The last layer has as many output nodes as regressed values. The topology of this baseline network are listed in Table \ref{tab:table_NetworkParameters}.
When we use this network for regression of the plane parameters, it is agnostic about the body region for which the planes’ parameters need to be calculated.

As in Martin et al.\cite{martin_vicario_automatic_2020}, this information was provided by selecting the correct individual network. We want to compare the performance of this base network with two extended versions which also use the additional class information. In the first version, we encode the information as a $\left(N \times 1\right)$ one-hot encoded tensor which bypasses the convolutional layers and is concatenated to the output of the last convolutional layer and fed into the first fully connected layer (Figure \ref{fig:Class_Network}). The idea behind this structure is that the convolutional layers are strictly limited to feature extraction and with processing information from more volumes of multiple anatomies, they can calculate more meaningful and generalizable features. Besides, the fully connected layers may benefit from this additional information.

Lastly, we investigate a multi-head approach with a shared convolutional feature extraction but individual fully connected regression heads for each anatomical region (Figure \ref{fig:Multihead_Network}) \cite{Ruder2017}. During inference the knowledge about the body region is used to select the head and output nodes that correspond to the given body region. During back-propagation, the error gradients for all other body regions are set to zero. Thus only parameters within the fully connected layers belonging to the selected body region and those within the convolutional layers are updated.

\begin{table}[]
\centering
\caption{Structure and parameter layout of the baseline network.}
\label{tab:table_NetworkParameters}
\begin{tabular}{@{}llll@{}}
\toprule
Block & Input resolution & \# Input channels & \# Output channels \\ 
\midrule
CNN1  & $72 \times 72 \times 72$     & $1$                 & $8$                  \\
CNN2  & $31 \times 37 \times 31$     & $8$                 & $16$                 \\
CNN3  & $16 \times 19 \times 16$     & $16$                & $32$                 \\
CNN4  & $8 \times 10 \times 8$       & $32$                & 64                 \\
CNN5  & $4 \times 5 \times 4$        & $64$                & $228$                \\ 
\midrule
FC1   & $1 \times 1 \times 1$        & $10240$             & $1300$               \\
FC2   & $1 \times 1 \times 1$        & $1300$              & $50$                 \\
FC3   & $1 \times 1 \times 1$        & $50$                & \# parameters      \\ 
\bottomrule
\end{tabular}

\end{table}

\subsection{Augmentation and Value Normalization}
During training, online augmentation of the volumes is employed. The spatial augmentation includes random rotation within the interval $\left[-45,\,45\right]^{\circ}$, random spatial scaling of the volume by a factor in the range $\left[0.95,\, 1.05\right]$, translation by $\left[-12, 12 \right]\,\si{mm}$, center cropping, and sub-sampling. All aforementioned augmentations were applied with a probability of $0.5$ and were sampled uniformly from the respectively given range. Additionally, mirroring in x-direction is added with a probability of $0.5$ which allows simulating left-right handedness of the volume. These spatial operations are composed by combining their representation by homogeneous matrices to a single composite matrix. The homogeneous transform matrix is given by
\begin{equation}
    T_m= T_r T_s T_t T_R
\end{equation}
where $T_r$, $T_s$, $T_t$, and $T_R$ represent respectively the  sub-sampling, scaling, translate and rotation homogeneous matrices. 
This way of implementation helps to speed up the calculation and reduces the number of performed interpolations to one.

Thereafter, also an intensity augmentation is implemented which simulates that the Hounsfield Unit (\si{HU}) values of mobile C-Arm devices are generally not as well calibrated as those of CT systems. Thereto, the value of $1000\,\si{HU}$ is added to the interpolated \si{HU} values, and the result is multiplied by a factor uniformly sampled from the range $\left[0.95, 1.05\right]$.
For normalization, a windowing function $w(x)$ is applied after clipping the volume intensity values to the range of $\left[-490, 1040\right]\,\si{HU}$ and rescaling it to $\left[0, 1\right]$.
The resulting intensity value before applying the windowing function is given by 
\begin{equation}
c(x)=\begin{cases}
0 & \text{if $x < \mathrm{min}$,} \\
\frac{f(x+1000)-\mathrm{min}}{\mathrm{max}-\mathrm{}{min}} & \text{if $ \mathrm{min} < x < \mathrm{max}$}, \\
1 & \text{if $x > \mathrm{max}$.}
\end{cases}
\end{equation}
where $f$ represents the random factor. The windowing function is defined as 
 \begin{equation}
w(x) = \frac{1}{(1+e^{g(0.5-x)})}    
\end{equation}
with a minimum and maximum value dependent gain factor. The gain factor is given by 
\begin{equation}
    g=\log \left ( \frac{1-y}{y} \right)/0.4
\end{equation} where $y= 0.02(\mathrm{max}- \mathrm{min})$. In contrast to min-max normalization, it reduces the signal variance of metal and air which typically contains little to no information about the plane’s parameters. 

\subsection{Post-processing of Regressed Values}
\label{subsec:postprocessing}

In Martin et al.\cite{martin_vicario_automatic_2020} it was shown that a combined regression of the parameters of the three planes is beneficial compared to train separate networks for each plane. So the accuracy can be improved when the planes are redundantly  regressed. In the same publication, it was also shown that the training does not benefit from an additional orthogonality constraint on the regressed values. Therefore, we decided to regress the parameters of the planes in all the presented architectures decoupled and adjust them afterwards algorithmically.

As presented in Martin et al.\cite{martin_vicario_automatic_2020}, the axial plane is the most accurately regressed in the anatomical regions. Therefore, it is taken as reference plane for other planes. That means, that the in-plane rotation of the coronal and the sagittal plane is corrected such that the intersection of the axial plane at these planes is at $0^{\circ}$. Thereafter, in cases in which the planes are orthogonal to each other, the normal direction of the sagittal plane is adjusted to be orthogonal to axial and coronal planes.

\section{Experiments}
\label{sec:experiments}
\subsection{Data Sets}
\begin{figure}[t]
    \centering 
    \caption{Representation of the 3D definition of axial (red), coronal (blue), and sagittal (green) standard planes in the calcaneus, ankle, knee, and wrist.}
    \label{fig:SP}    
    \includegraphics[width=0.98\textwidth]{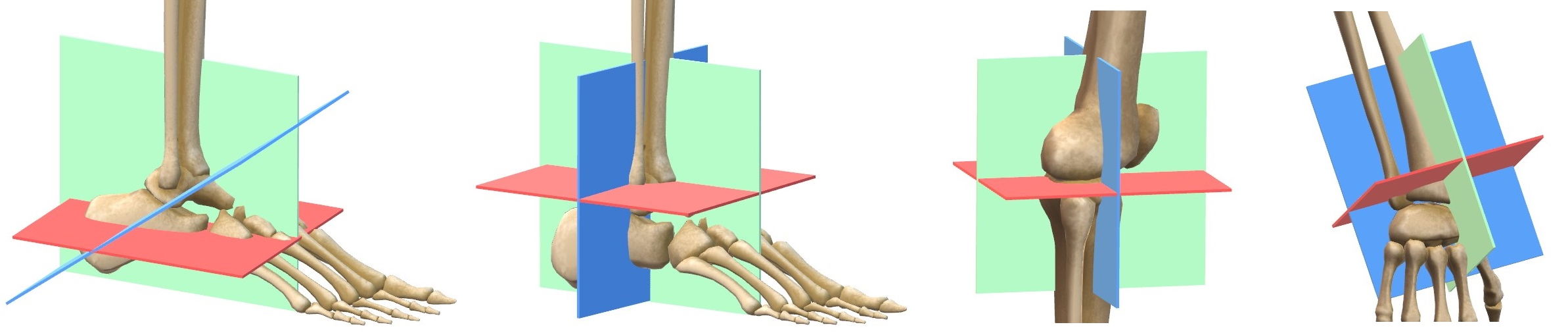}
\end{figure}

Our data set consists of 160 volumes of the calcaneus region, 220 volumes of the ankle region, 274 volumes of the knee, and 250 volumes of the wrist. All volumes were acquired with a mobile C-arm system Cios Spin from Siemens Healthineers and reconstructed offline with Feldkamp-David-Kress algorithm using parameters equal to the product standard settings. The volumes have a uniform resolution of $512^3$ voxels and a field of view of $(160 \si{mm})^3$. They were partly acquired after an orthopedic surgery for assessing the surgical result and partly were acquired from cadavers which were prepared for surgical training. The cadaver data sets were typically scanned twice: once without any metal and once with metal objects put on the surface of the cadaver. We also obtained volumes of cadavers with various metal implants acquired during surgical training. The exact distribution of the data sets is listed in Table~\ref{tab:table_datasets}. All available volumes were included in the data set, without any constraint on the positioning of the body part of interest. The volumes were corrected for wrong patient position description according to the DICOM meta information.
For each body region 5 data splits were created, taking care that volumes of the same
patient belong to the same subset and that the distribution of the data set's origin is approximately the same as in the total data set. For all volumes, standard planes were defined according to the clinical definition provided in Grützner\cite{grutzner_rontgenhelfer_2004}. Sketches of the planes are displayed in Figure \ref{fig:SP}.

For the ankle, knee, and wrist volumes axial, coronal, and sagittal MPRs, for the calcaneus data sets axial, sagittal, and semi-coronal planes were annotated. This was done by a medical engineer after five hours of training using a syngo XWorkplace VD20 which was modified to store the plane description. Axial, sagittal, and coronal MPRs were adjusted with coupled MPRs. The semi-coronal plane was adjusted thereafter with decoupled planes. The annotation validity was verified by an expert physician and additionally by a senior medical engineer.

\begin{table}[]
\centering
\caption{Number, origin, and realism with respect to metallic objects of the volumes.}
\label{tab:table_datasets}
\begin{tabular}{lllllll} 
  \toprule
               & \multicolumn{3}{c}{Cadaver}                                                                                                                                             && Clinical                                                 & Total \\ \cmidrule{2-4} \cmidrule{6-6}
               & \begin{tabular}[c]{@{}l@{}}Metal \\ implants\end{tabular} & \begin{tabular}[c]{@{}l@{}}Metal\\ outside\end{tabular} & \begin{tabular}[c]{@{}l@{}}No\\ Metal\end{tabular} & & \begin{tabular}[c]{@{}l@{}}Metal\\ implants\end{tabular} &       \\ \midrule
Calcaneus      & $9$                                                         & $63$                                                      & $62$                                                 & &$26$                                                       & $160$   \\
Ankle          & $36$                                                        & $61$                                                      & $56$                                                & & $67$                                                       & $220$   \\
Knee & $65$                                                        & $68$                                                      & $70$                                                 && $71$                                                       & $274$   \\
Wrist  & $0$                                                         & $101$                                                     & $102$                                               & & $46$                                                       & $249$   \\ \bottomrule
\end{tabular}

\end{table}

\subsection{Performance Metric}
\label{subsec:performancemetric}
As an evaluation metric to compare the performance of the networks, we use a weighted average over the individual error values of the three regressed planes
\begin{equation}
\label{equ_weight}
	p =\frac{1}{\#\mathrm{planes}} \sum_{j \in \mathrm{planes}}0.2d_j + 0.6\epsilon_{n,j} + 0.2\epsilon_{i,j} .
\end{equation}
$d_j$ denotes the mean error of the absolute translation of the center in direction of the $j^{th}$ plane's normal. $\epsilon_{n,j}$ is the deviation of the normal vectors $e_w$, and $\epsilon_{i,j}$ is the in-plane rotation error calculated as mean difference angle of $e_u$ and $e_v$, after projecting the directions on the plane defined by the annotation. The different weights in \eqref{equ_weight} were chosen heuristically and reflect that the normal has the most complex effect on the result. For this normal to be corrected, out-of-plane rotations would be necessary, whereas in-plane rotation and plane translation are easy-to-fix components.

In the results tables below, the mean and the standard deviation of the median prediction errors of the folds is represented. 

\subsection{Study Design}
Before investigating a combined regression network for multiple anatomies, some further experiments were carried out to evaluate the performance of the baseline network. We have seen in Section \ref{sec:methods} that there are several possibilities to parameterize rotations. In addition to Martin et al. \cite{martin_vicario_automatic_2020} the $\textrm{6D}_{xz}$ method was introduced taking into account that the main contribution to the performance metric comes from angular deviation of the normals. Therefore, as first experiment the comparison of the representation with Euler angles, quaternions, $\textrm{6D}_{xy}$, and $\textrm{6D}_{xz}$ is performed for the four body regions.

The best performing representation is used in the subsequent experiments.
First, the influence of the post-processing of the regressed angles is evaluated. For that, $\epsilon_n$ and $\epsilon_i$ are calculated with and without post-processing and their values are compared.

In Martin et al.\cite{martin_vicario_automatic_2020} the question was kept open, whether better results can be expected with more data samples. Since the number of available volumes is fixed, we incrementally reduce the number of volumes used for training. For this, the training for the different body regions is repeated using $100\,\%$, $80\,\%$, $60\,\%$, and $40\,\%$ of the volumes in the training split, while keeping the test volumes unchanged.

Following the evaluation of the baseline model, different experiments were carried out to evaluate the performance of the use of a single model for all the body regions. First, we trained a single network for all body regions without providing any further class information. Second, we extend this architecture by encoding the anatomical region information as a one-hot encoded vector and concatenating it with the output of the convolutional layers (Fig. \ref{fig:Class_Network}). Lastly, a multi-head architecture (Fig. \ref{fig:Multihead_Network}) is used where all body regions share the convolutional feature extraction layers but are individually processed in separate regression heads consisting on three fully connected layers for each anatomical region. In order to overcome the imbalance between the different classes, the volumes were randomly over-sampled from the minority classes with a weight given by the number of volumes from a given class. To gain a better understanding of the influence and benefits of the additional anatomy information in the case of the first extended model, this additional information is corrupted to varying degrees during model inference. For this purpose, all nodes of the one-hot-tensor are set to the same expected diffuse probability of either $0.0$, $0.5$, or $1.0$.

\subsection{Implementation}
The models are implemented in PyTorch $\left(v.1.5.1\right)$ and trained on Windows 10 systems with $32\,\si{GB}$ RAM and $8\,\si{GB}$ NVIDIA RTX 2070S. The weights are initialized by the He et al. method \cite{he_kaiming_deep_2015}. The network is trained by a mini-batch gradient descent optimizer with momentum. For optimization of the network parameters, the mean squared error between model prediction and ground truth was calculated at each output node. The total number of epochs was set to $400$, verifying training convergence of all model variants. For the selection of the learning rate, learning rate decay, step size, momentum, and batch size, a hyper-parameter optimization using random sampling of the search space was performed. For that purpose, one fold was used and individual hyper-parameter optimizations were performed for the different rotation descriptions in the baseline network. In Table \ref{tab:Search_space} the search space for each hyper-parameter evaluated as well as the sampling value for the $6D_{xy}$ representation are listed. This method results in an offset of typically $0.1$ and maximum $0.4$ score points. 

\begin{table}[h]
\centering
\caption{Search space hyper-parameters, sampling distribution and best configuration for the plane regression task as result of random search hyper-parameter optimization. }
\label{tab:Search_space} 

\begin{tabular}{@{\extracolsep{0.5pt}}lcc@{}}
\toprule
Hyper-parameter & Sampling distribution & Sampling value\\
\midrule
Learning rate    & $ s \sim \log \mathcal{U}(0.0001, 0.01)$ & $0.00164$\\ 
Learning rate decay & $ s \sim \log \mathcal{U}(0.2, 0.9)$ & $0.27291$\\
Learning rate decay step & $ s \sim  \mathcal{U}(20, 80) $ & $75$\\
Momentum  & $  s \sim \log \mathcal{U}(0.5, 0.99) $  & $0.957437$\\ 
\midrule
Batch size &   $ s \sim  \mathcal{U}(5, 12) $  & $9$\\ 
\bottomrule
\end{tabular}
\end{table}

\section{Results}
\label{sec:results}
As can be observed in Table \ref{tab:results_rotation}, the evaluation of the different rotation representations in the base model shows that the 6D method outperforms the Euler and quaternion representations in all the body regions except the knee. For this region similar performance to the best representation, the Euler angles, is reached. Among the 6D methods, no significant difference in performance between $\textrm{6D}_{xz}$ and $\textrm{6D}_{xy}$ can be observed. Thus, using the normal in the directly obtained values and consequently also in the cost function does not generally improve the quality of the planes parameter regression. In two body regions we could observe a small reduction in the mean error of the estimated normals, whereas an error increase was registered for the other two regions. In all cases the in-plane rotation performance got significantly worse. The position estimation of the planes was approximately the same for both representations. Due to these reasons, the $\textrm{6D}_{xy}$ variant was chosen for the remaining experiments. When using sine and cosine representations of the Euler angles instead of the raw angle values shows superior performance over the quaternion representation for the estimation of the plane normal. Looking at the performance score that weights all metrics (Subsection \ref{subsec:performancemetric}), the Euler angles show better results in three body regions compared to the quaternions.

\begin{table}[h]
\centering
\caption{Summarized results of evaluation of Euler angles, quaternions, $\textrm{6D}_{xy}$, and $\textrm{6D}_{xz}$ rotation representations in standard plane regression of calcaneus, upper ankle, knee, and wrist regions.}
\label{tab:results_rotation}
\begin{tabular}{@{\extracolsep{2pt}}lccccc@{}}
\toprule
\multicolumn{2}{l}{}                    & d (mm)        & $\varepsilon_n (^{\circ})$ & $\varepsilon_i (^{\circ})$ & Score          \\ \midrule
\multicolumn{6}{l}{Calcaneus}                \\
\multicolumn{2}{c}{Euler}               & 14.39 ± 1.64  & 8.93 ±   1.60       & 9.99 ± 0.75          & 10.23 ± 1.11   \\
\multicolumn{2}{c}{Quat.}               & 9.93 ± 2.53   & 9.96 ±   1.75       & 9.57 ± 1.52          & 9.87 ± 1.65    \\
\multicolumn{2}{c}{$\textrm{6D}_{xy}$}  & 9.94 ±   1.92 & \textbf{8.08 ± 0.38}         & \textbf{8.09 ± 0.45}          & \textbf{8.46 ± 0.63}    \\
\multicolumn{2}{c}{$\textrm{6D}_{xz}$}  & \textbf{9.31 ±   1.10} & 8.23 ±   0.69       & 9.42 ±   1.02        & 8.68 ±   0.65  \\ \midrule
\multicolumn{6}{l}{Ankle}              \\
\multicolumn{2}{c}{Euler}               & 7.78 ±   0.36 & 6.98 ± 0.77         & 7.52 ± 0.76          & 7.25 ± 0.66    \\
\multicolumn{2}{c}{Quat.}               & \textbf{5.00 ±   0.09} & 8.16 ±   0.79       & 8.31 ±   0.71        & 7.56 ±   0.63  \\
\multicolumn{2}{c}{$\textrm{6D}_{xy}$}  & 5.43 ±   0.25 & 6.61 ± 0.34         & \textbf{6.37 ± 0.31}          & 6.32 ± 0.25    \\
\multicolumn{2}{c}{$\textrm{6D}_{xz}$}  & 5.41 ±   0.49 & \textbf{6.17 ±   0.78}       & 7.32 ±   1.06        & \textbf{6.25 ±   0.65}  \\ \midrule
\multicolumn{6}{l}{Knee}      \\
\multicolumn{2}{c}{Euler}               & \textbf{6.81 ±   0.65} & \textbf{6.59 ±   1.05}       & 7.36 ±   1.54        & \textbf{6.79 ±   0.96}  \\
\multicolumn{2}{c}{Quat.}               & 6.82 ±   0.72 & 9.45 ±   0.67       & 10.54 ±   1.22       & 9.15 ±   0.59  \\
\multicolumn{2}{c}{$\textrm{6D}_{xy}$}  & \textbf{6.81 ±   0.47} & 6.71 ±   0.63       & \textbf{7.07 ±   0.95}        & 6.80 ±   0.55  \\
\multicolumn{2}{c}{$\textrm{6D}_{xz}$}  & 7.15 ±   1.16 & 7.19 ±   0.52       & 8.22 ±   0.62        & 7.39 ±   0.49  \\ \midrule
\multicolumn{6}{l}{Wrist}   \\
\multicolumn{2}{c}{Euler}               & 7.45 ±   1.00 & 8.35 ±   1.66       & 9.82 ±   1.38        & 8.48 ±   1.31  \\
\multicolumn{2}{c}{Quat.}               & 8.46 ±   1.93 & 11.31 ±   1.87      & 13.47 ±   2.48       & 11.22 ±   1.83 \\
\multicolumn{2}{c}{$\textrm{6D}_{xy}$}  & 7.27 ±   1.08 & 7.74 ±   1.14       & \textbf{8.72 ±   0.64}        & 7.85 ± 0.94    \\
\multicolumn{2}{c}{$\textrm{6D}_{xz}$}  & \textbf{7.21 ±   1.02} & \textbf{7.21 ±   1.06}       & 9.37 ±  1.08        & \textbf{7.64 ±   0.84}  \\ \bottomrule
\end{tabular}
\end{table}

The analysis of the influence on the post-processing to the single parts of the score for $\textrm{6D}_{xy}$ representation (Table \ref{tab:results_coupled}) shows that the post-processing helps to significantly improve $\epsilon_n$ as well as $\epsilon_i$ by up to $1.89^{\circ}$. As the translation remains untouched by the post-processing, no changes can be observed for the translation error $d$.

\begin{table}[h]
\centering
\caption{Comparison of the errors directly obtained by the network (Regressed) and after post-processing ensuring orthogonality of respective planes (Post-proc.) using the {$\textrm{6D}_{xy}$} rotation representation.}
\label{tab:results_coupled}
\begin{tabular}{@{\extracolsep{0.5pt}}lcccc@{}}
\toprule
 & {d (\si{mm})} & {$\varepsilon_n (^{\circ})$} & {$\varepsilon_i (^{\circ})$  } & {Score} \\ \midrule
Calcaneus                                                                                                                                              \\
\:Regressed & \textbf{9.94 ±   1.92} & 8.77 ± 0.60 & 8.34 ± 0.44  & 8.92 ± 0.51               \\
\:Post-proc.  & \textbf{9.94 ± 1.92} & \textbf{8.08 ± 0.38} & \textbf{8.09 ± 0.45} & \textbf{8.46 ± 0.63}             \\ \midrule
Ankle                                                                                                                                                   \\
\:Regressed & \textbf{5.43 ±   0.25} & 7.11 ± 0.48  & 6.58 ± 0.29 & 6.70 ± 0.35               \\
\:Post-proc.   & \textbf{5.43 ± 0.25} & \textbf{6.61 ± 0.34} & \textbf{6.37 ± 0.31} & \textbf{6.32 ± 0.25}             \\ \midrule
Knee                                                                                                                                          \\
\:Regressed & \textbf{6.81 ± 0.47}              & 8.60 ± 0.98                             & 8.45 ± 0.63                              & 8.22 ± 0.71               \\
\:Post-proc.   & \textbf{6.81 ± 0.47}              & \textbf{6.71 ± 0.63}                           & \textbf{7.07 ± 0.95}                            & \textbf{6.80 ± 0.55}             \\ \midrule
Wrist                                                                                                                                           \\
\:Regressed & \textbf{7.27 ± 1.08}              & 8.76 ± 1.19                             & 8.84 ± 0.94                              & 8.48 ± 1.09               \\
\:Post-proc.   & \textbf{7.27 ± 1.08}              & \textbf{7.74 ± 1.14}                           & \textbf{8.72 ± 0.64}                            & \textbf{7.85 ± 0.94}\\ \bottomrule             
\end{tabular}
\end{table}

The performance analysis of the baseline model upon reduced amounts of training data (Table \ref{tab:ablation}) reveals that in the ankle body region 174 volumes are sufficient to find good results. For the other body regions the numbers of provided volumes should be increased to obtain the best possible results. Compared to the ankle, the other regions show a larger variance in shape and joint angulation and thus more training data is needed to capture all different shapes. It can be observed that calcaneus, knee, and wrist regions all show a similar performance characteristics at reduced amounts of training data.

\begin{table}
\centering
\caption{Summarized results of evaluation of Euler angles, quaternions, $\textrm{6D}_{xy}$, and $\textrm{6D}_{xz}$ rotation representations in standard plane regression with (Post-proc.) and without (Regressed) post-processing.}
\label{tab:results_rotation}
\begin{tabular}{@{\extracolsep{0.5pt}}lcccc@{}}
\toprule
 & {Calcaneus} & {Ankle} & {Knee } & {Wrist} \\ \midrule
 Euler
 \\
\:Regressed & 10.44 ±   1.17 & 7.42 ± 0.77 & \textbf{6.68 ± 0.85}  & 8.30 ± 1.24               \\
\:Post-proc.  & 10.23 ± 1.11 & 7.25 ± 0.66 & 6.79 ± 0.96 & 8.48 ± 1.31            \\ \midrule
Quat. 
\\
\:Regressed & 9.88 ±   1.65 & 7.70 ± 0.67  & 9.01 ± 0.65 & 11.38 ± 2.00               \\
\:Post-proc.   & 9.87 ± 1.65 & 7.56 ± 0.63 & 9.15 ± 0.59 & 11.22 ± 1.83            \\ \midrule
$\mathrm{6D}_{xy}$                                                                     \\
\:Regressed & 8.92 ± 0.51             & 6.70 ± 0.35                             & 8.22 ± 0.71                              & 8.48 ± 1.09               \\
\:Post-proc.   & \textbf{8.46 ± 0.63}              & 6.32 ± 0.25                          & 6.80 ± 0.55                           & 7.85 ± 0.94            \\ \midrule
$\mathrm{6D}_{xz}$                                                                                                                                       \\
\:Regressed & 8.71 ± 0.66             & 6.58 ± 0.64                             & 7.59 ± 0.60                              & 8.05 ± 1.07               \\
\:Post-proc.   & 8.68 ± 0.65             & \textbf{6.25 ± 0.65}                           & 7.39 ± 0.49                            & \textbf{7.64 ± 0.84}\\ \bottomrule             
\end{tabular}

\end{table}

\begin{table}[h]
\centering
\caption{Summarized results of the different  networks including the use of multiple models (a model for each anatomy), of a model for training all the anatomies without class information, of a model trained with class information, and the multi-head model.}
\label{tab:results_models}
\begin{tabular}{@{\extracolsep{0.5pt}}lcccc@{}}
\hline
   & {d (\si{mm})} & {$\varepsilon_n (^{\circ})$} & {$\varepsilon_i (^{\circ})$} & {Score} \\ \hline
Calcaneus                                                                                                                                               \\
\:Multiple   & 9.94 ± 1.92                & \textbf{8.08 ± 0.38}                             & \textbf{8.09 ± 0.45}                              & \textbf{8.46 ± 0.63}               \\
\:w/o class  & 9.17 ± 0.64                & 9.18 ± 1.21                             & 8.87 ± 1.34                              & 9.12 ± 1.03               \\
\:w/ class   & 9.38 ± 1.30                & 9.90 ± 2.16                             & 10.31 ± 2.26                             & 9.88 ± 1.71               \\
\:Multi-head & \textbf{7.44 ± 0.31}                & 9.16 ± 1.80                             & 8.55 ± 0.88                              & 8.69 ± 1.23               \\ \hline
Ankle                                                                                                                                                  \\
Multiple   & 5.43 ± 0.25                & 6.61 ± 0.34                             & \textbf{6.37 ± 0.31}                              & 6.32 ± 0.25               \\
\:w/o class  & 6.34 ± 0.77                & 9.71 ± 1.98                             & 9.64 ± 1.90                              & 9.02 ± 1.59               \\
\:w/ class   & 5.55 ±   1.11              & 7.71 ± 1.14                             & 8.73 ± 1.68                              & 7.49 ± 0.97               \\
\:Multi-head & \textbf{4.47 ± 0.33}                & \textbf{6.08 ± 0.45}                             & 6.61 ± 0.65                              & \textbf{5.86 ± 0.40}               \\ \hline
Knee                                                                                                                                           \\
\:Multiple   & 6.81 ± 0.47                & 6.71 ± 0.63                             & 7.07 ± 0.95                              & 6.80 ± 0.55               \\
\:w/o class  & 6.71 ± 0.72                & 8.04 ± 0.58                             & 8.14 ± 1.10                              & 7.79 ± 0.53               \\
\:w/ class   & 6.34 ± 0.82                & 8.16 ± 1.35                             & 7.95 ± 1.21                              & 7.75 ± 1.11               \\
\:Multi-head & \textbf{5.62 ± 0.68}                & \textbf{6.70± 1.28}                              & \textbf{6.77± 0.81}                               & \textbf{6.49 ± 1.05}               \\ \hline
Wrist                                                                                                                                            \\
\:Multiple   & 7.27 ± 1.08                & \textbf{7.74 ± 1.14}                             & \textbf{8.72 ± 0.64}                              & \textbf{7.85 ± 0.94}               \\
\:w/o class  & \textbf{6.42 ± 0.75}                & 10.34 ± 2.82                            & 10.52 ± 2.04                             & 9.59 ± 2.15               \\
\:w/ class   & 6.46 ± 1.06                & 10.52 ± 2.33                            & 10.00 ± 1.63                             & 9.61 ± 1.77               \\
\:Multi-head & 7.03 ± 1.16                & 10.50 ± 1.73                            & 11.15 ± 1.29                             & 9.93 ± 1.41               \\ \hline
\end{tabular}
\end{table}

The comparison of the multi-head networks (Table \ref{tab:results_models}) shows that a combined network which jointly estimates the parameters of the planes for different body region can in two cases improve the accuracy of the planes positions. However, for the angle regression task, this network variant yields inferior results. As the angular errors have a higher impact on the score, the overall performance is inferior. 

The same holds true for the network architecture where all layers are shared by all body regions and additionally the body region is provided to the fully connected layers. This architecture has approximately the same score as the network without the additional class information. Analyzing the influence of information about the body region shows that the current network and training configuration is not suitable for incorporating and interpreting this additional information.

\begin{table}[t]
\caption{Performance comparison for different provision levels of anatomical class information. The network is provided once with true information about the volume class (one-hot) and for three different diffuse priors where equal probabilities are assigned at each class label node ({$0$, $0.5$, $1$}).}
\label{tab:usage_class_information}
\centering
\begin{tabular}{@{}lcccc@{}}
\toprule
 & Calcaneus & Ankle  & Knee    & Wrist               \\ 
\midrule
True label  & \textbf{9.88 ± 1.71}  & 7.49 ± 0.97          & 7.75 ± 1.11           & 9.61 ± 1.77 \\
0.0  & 11.18 ± 1.08          & 7.45 ± 0.97          & 8.15 ± 1.00           & 9.76 ± 1.53         \\
0.5  & 10.73 ± 1.11          & \textbf{6.67 ± 1.48} & 8.28 ± 1.35           & \textbf{9.33 ± 1.09}\\ 
1.0  & 10.53 ± 0.67          & 7.95 ± 0.66          &  \textbf{7.68 ± 1.11} & 9.38 ± 1.46         \\ 
\bottomrule
\end{tabular}
\end{table}

The multi-head network achieves the best performance score for 2 out of 4 body regions. For the calcaneus region the single task network and the multi-head network have about the same performance, with their mean performance score and rotation errors lying in each others range of standard deviation. Only for the wrist body region the angle errors and thus also the score is significantly worse compared to the single task network. For this region, the multi-head network has achieved the worst values compared to all MTL network variants. 

\begin{figure}[h]
    \centering    
    \caption{Individual distribution of plane and distance errors per anatomy obtained by the multi-head network (\ref{subsec:performancemetric}).} \label{fig:error_distribution}
    \includegraphics[width=0.48\textwidth]{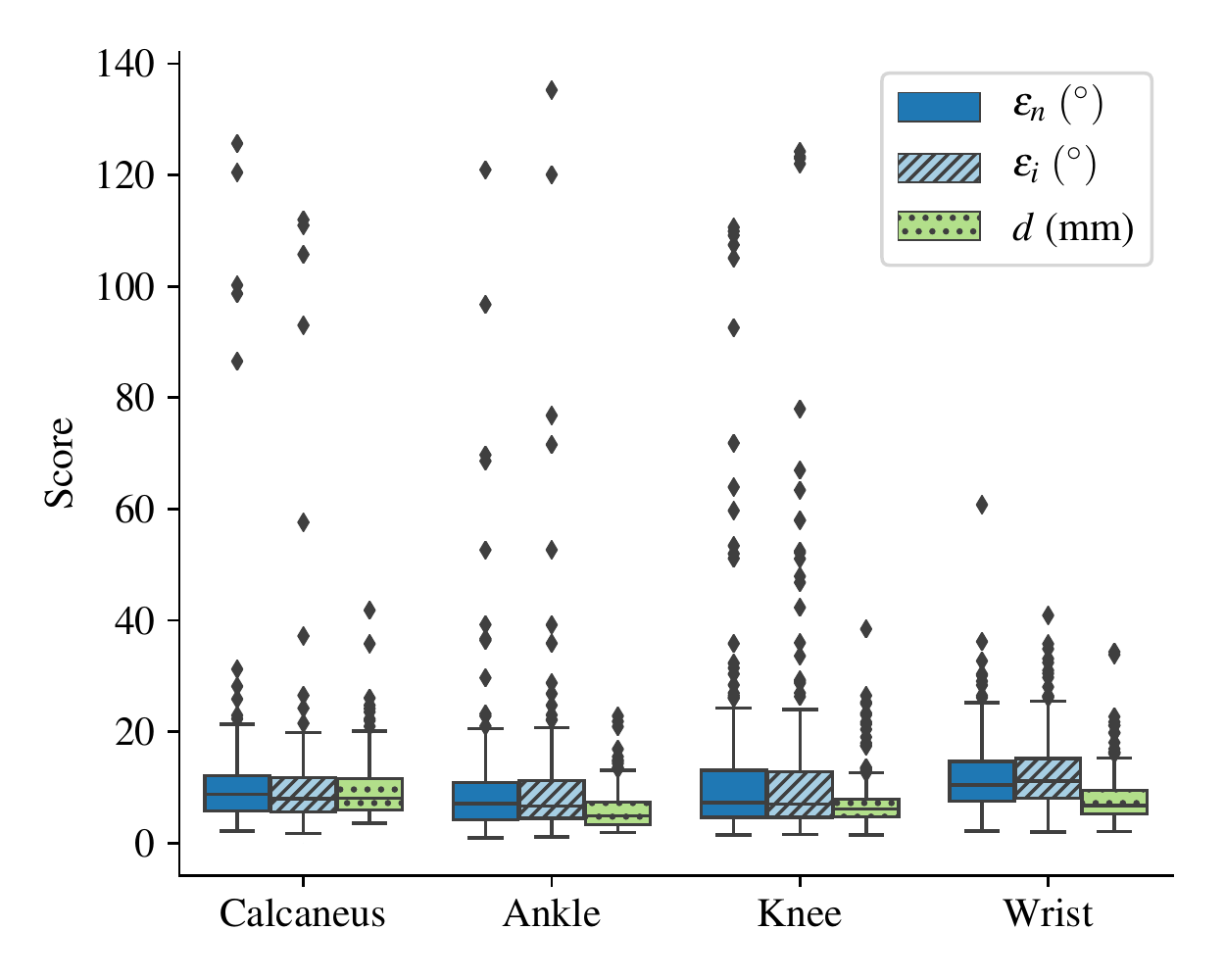}
\end{figure}

Across all experiments we could see that the estimation of the position can be improved by the MTL approaches (Table \ref{tab:results_models}). However, the angle estimation for both the normals and the in-plane rotation -- which typically have a higher nominal error compared to the position (Figure \ref{fig:error_distribution}) -- do not benefit from the MTL approach. 

For a better understanding of this result, we compared the volumes contributing to the $10\,\%$ best scoring results to those contributing to the $10\,\%$ worst scoring results. The presence of metallic objects like screws or plates could not be observed as a source for these errors. Likewise, we could preclude that the regression error is higher in those volumes where only a portion of the relevant anatomy is represented. For these problematic cases, the algorithm is quite robust.
However, in these volumes, we realized that the patient positioning was done in a different way in comparison to the standard, e.g. prone or left instead of supine, or focus on the proximal femur instead of the tibial head.

Since we constrained the augmentation pipeline by purpose not to fully cover this flips and rotations, more training data needs to be added to handle this.

\begin{figure}[t]
    \centering
    \caption{
     Example of automatic plane regression results by the multi-head network for the clinical wrist data set.}
    \includegraphics[width=0.60\textwidth]{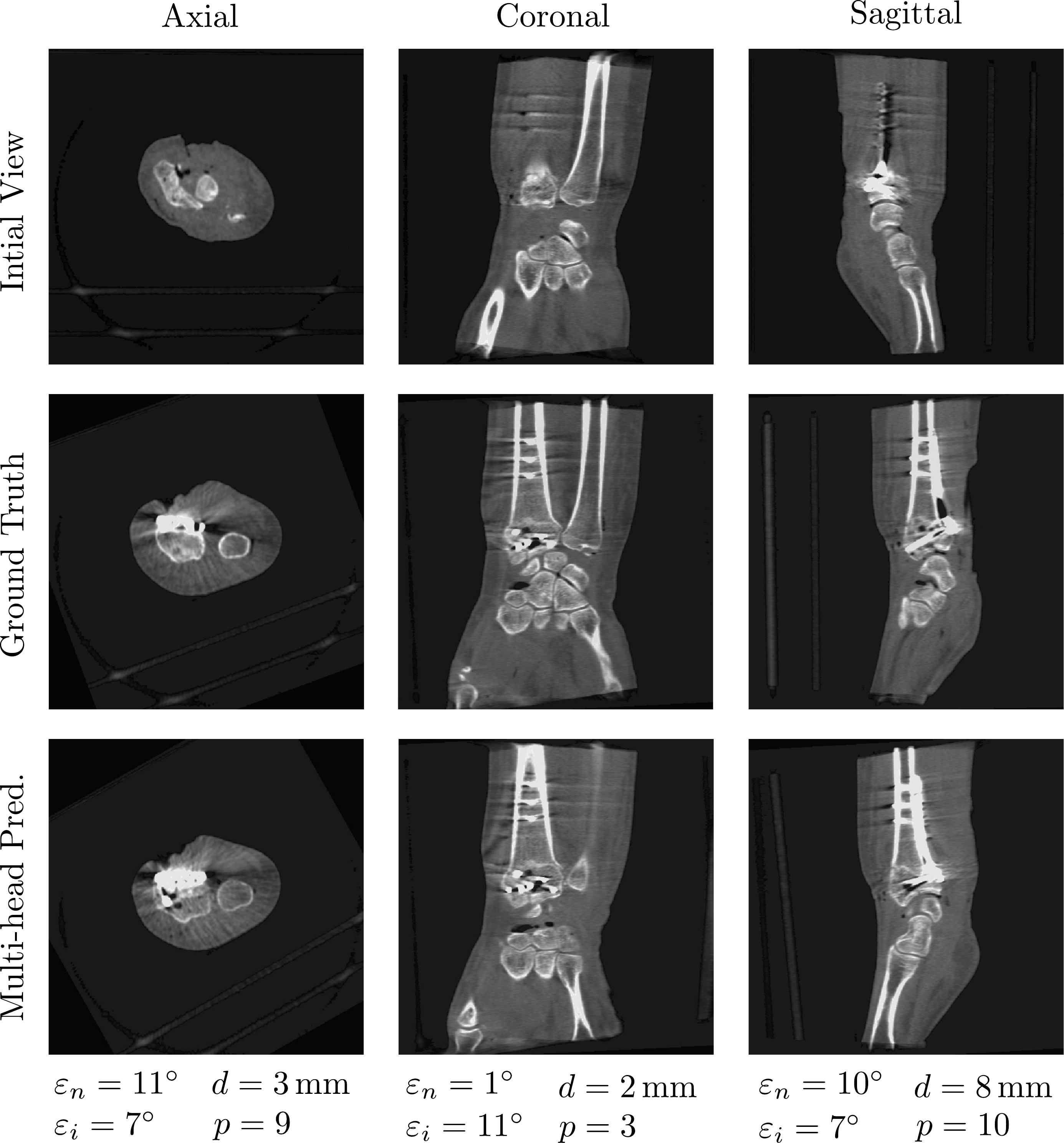}\label{fig:planes_radius}
\end{figure}

\begin{figure}[t]
    \centering
    \caption{
    Example of automatic plane regression results by the multi-head network for the clinical calcaneus data set.}
    \label{fig:planes_calcaneus}
    \includegraphics[width=0.60\textwidth]{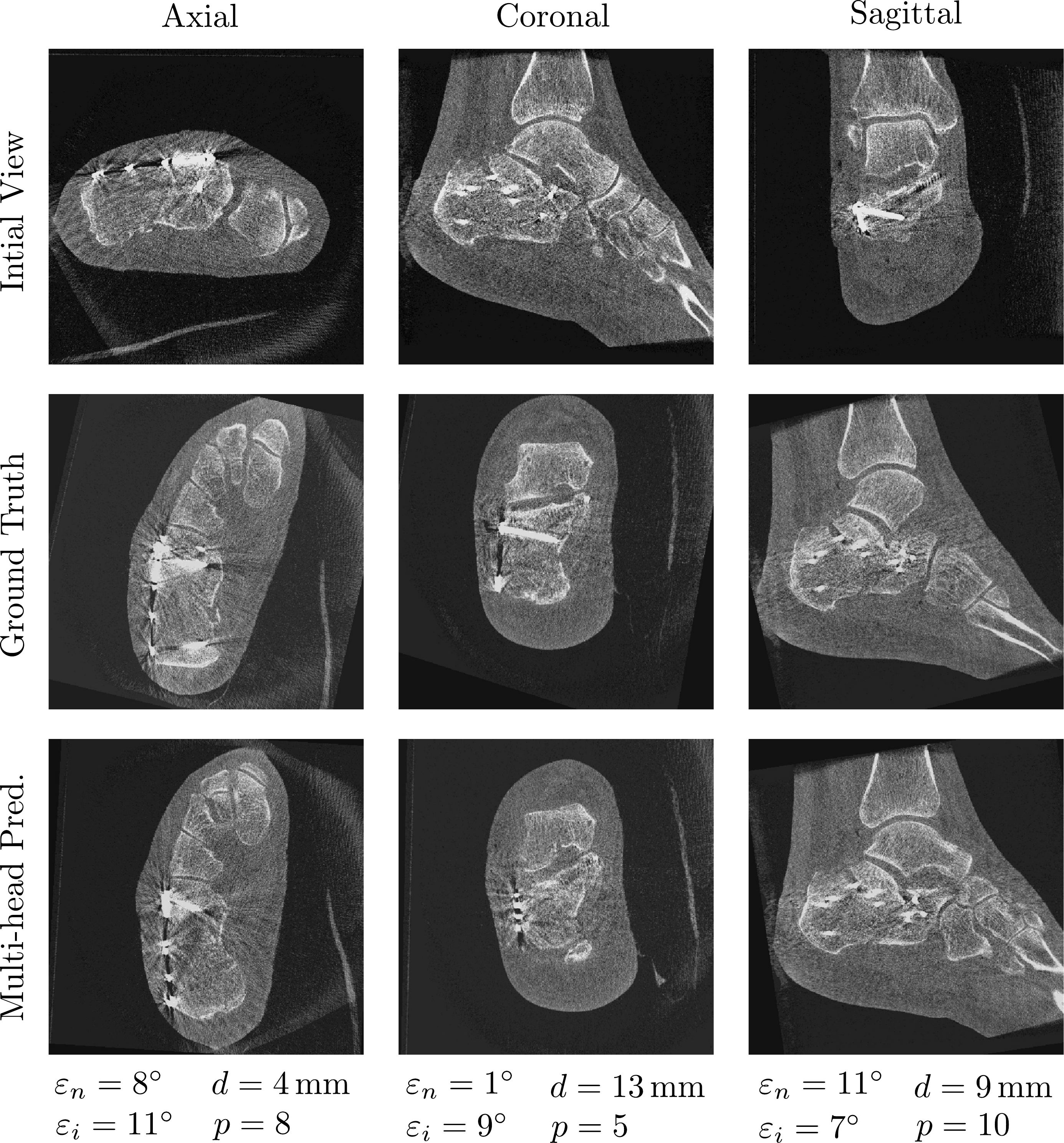}
\end{figure}

\begin{figure}[t]
    \centering
    \caption{
    Failure example of the automatic plane regression by the multi-head network for the clinical calcaneus data set.}
    \label{fig:calcaneus_bad}
    \includegraphics[width=0.60\textwidth]{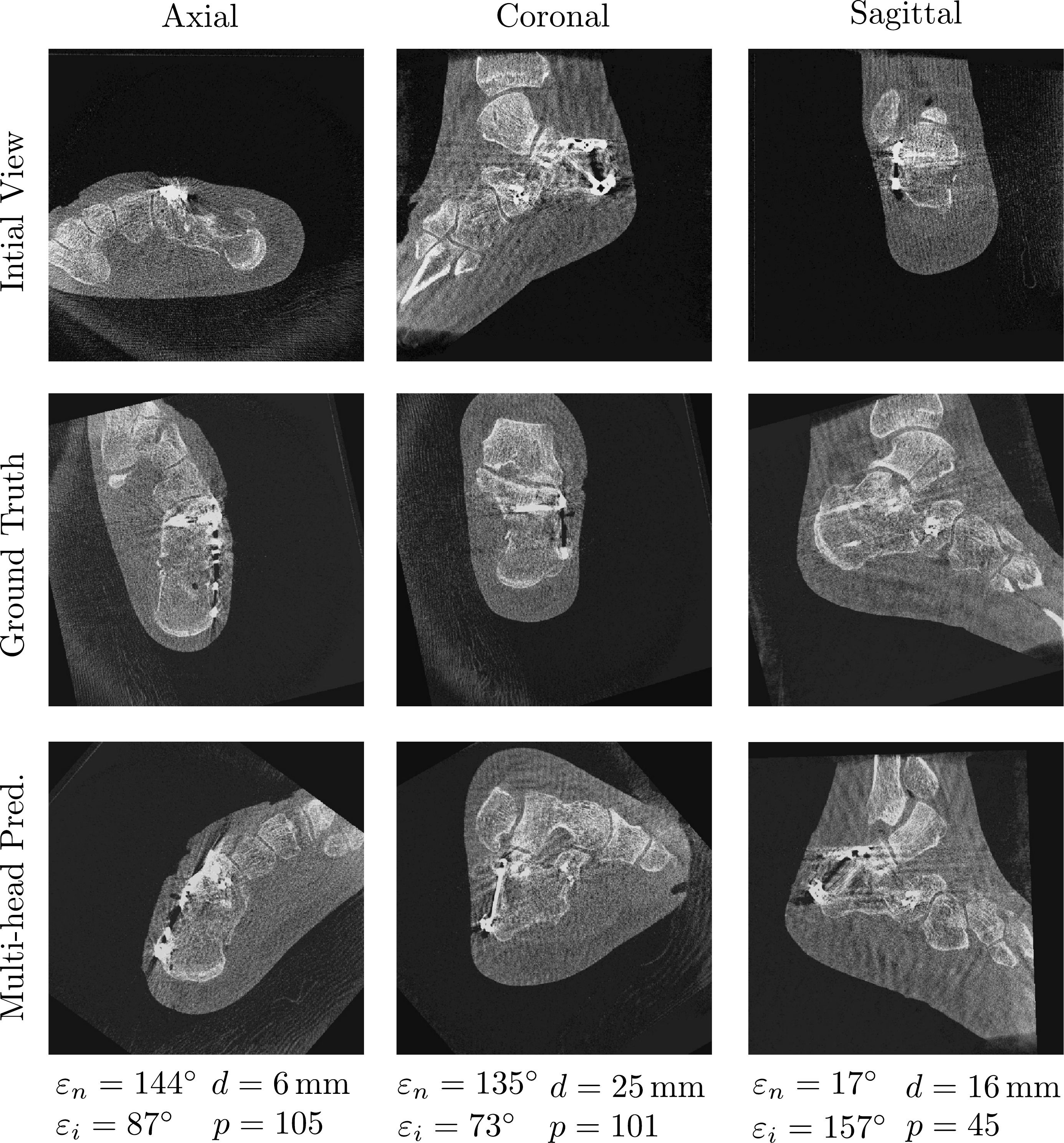}
\end{figure}

Figures \ref{fig:planes_radius}-\ref{fig:calcaneus_bad} show samples of the central planes through clinically acquired CBCT volumes and compare them to both the manually adjusted standard planes as well as the automatically inferred predictions by the multi-head network. In some cases, the algorithm could correct the in-plane rotation by $180^{\circ}$ (Figure \ref{fig:planes_radius}) or plane flips (Figure \ref{fig:planes_calcaneus}). However, the rotation by $90^{\circ}$ of the axial plane in Figure \ref{fig:calcaneus_bad} could lead to a very bad regression result.

\section{Discussion and Conclusion}
In this paper we investigate the regression of standard planes for four different body regions. The volumes for which the standard planes should be regressed are acquired with mobile C-arm devices and therefore have a limited field of view. Furthermore, there is no standardized relationship between the C-arm device and the body region of interest, which also means that the representation of the body region in the acquired volumes is not consistent. This also applies to the position of the body region in relation to the operating table. The target body regions are also in close proximity to flexible joints like knee, wrist, or ankle, which leads to great variability of the input data and thus to a significantly higher task complexity. 

Despite this complex setting, our proposed method yields encouraging results with low median errors for the regressed angles and also positions. The experiment results reveal that the single task networks already achieve a very good accuracy and that in this context MTL can only improve the results by a small margin. Only the multi-head approach could produce significantly better results. We argue that the used configuration of the fully connected layers in the combined network is not capable of learning an appropriate representation of data and task distribution, so that the addition of further information does not help to improve regression performance. By corrupting the anatomy information provided to the model with shared fully connected layers, the results did not get worse. This observation leads to the conclusion that the additional information has no influence on the network's decision-making. These shortcomings could be addressed by performing feature abstraction and combination in smaller consecutive steps, for example by adding intermediate fully connected layers. This reasoning is supported by the observation that only the additional parameter capacity of the multi-head approach has led to a performance increase. 

An important step to enhance the accuracy of the plane regression is to couple the planes as a post-processing step. We could show that this improves $\epsilon_n$ and $\epsilon_i$ by up to $1^{\circ}$. The improvement obtained by this post-processing method exceeds the one obtained by integration of the plane coupling in the training by a modified cost function as was studied in \cite{martin_vicario_automatic_2020}. Most of all, the post-processing guarantees the orthogonality of the coupled planes.

The job of regressing the planes parameters can be performed equally well for orthogonal and oblique planes. While for the specialized networks for the calcaneus body region with oblique planes were obtained, for the multi-head network the worst values were received for the wrist body region. For this body region the planes are orthogonal. While axial planes are typically well regressed, the overall score is deteriorated by the coronal and sagittal planes. The normals of these planes are typically less well defined and small rotations by a few degrees are hardly noticed.

The results show that good angle regression performance is obtained when the volumes are acquired with the body moderately aligned to the imaging system axes but fails when the body is rotated by more than $90^{\circ}$. For these cases, the applied augmentation pipeline does not help. Flips in y-direction were not covered by the augmentation. This was done by purpose, since in clinical practice a flip of a wrist upside down comes along with a modification of the configuration. In the case of the upper ankle or calcaneus, the upper ankle joint gets stretched more. Thus, applying the augmentation does not lead to clinically relevant data sets. Since at present stage additional clinical data is not available and their clinical acquisition is seldom, more cadaver data is needed to sufficiently represent those poses. This does also mean that the results presented in this work do not show the full potential of this approach. 

Kausch et al.\cite{kausch_toward_2020} has shown that the human performance in adjusting the planes highly depends on the target region. In regions with many well-defined landmarks and few anatomical variations, the inter-rater variance in the plane adjustment is low. However, in regions for which less reliable landmarks can be identified, this inter-rater variance is substantially higher. 

For the presented anatomies no such variance estimates are available yet. This imposes limitations on the interpretability of our results, since no well-defined reference values for clinically required error limits can serve as a standard. Although such a comparative analysis should be addressed in follow-up studies, we generally see promising results of our proposed method which fits well within the error bounds of related studies of anatomy with comparable complexity \cite{kausch_toward_2020}.

The benefit of the direct standard plane parameter regression is clearly the reduced amount of annotation data per data set. Costly annotations of landmarks or even segmentation of bones can be omitted and are replaced by comparably cheap adjustments of the standard planes. Also the implementation of specific rules per body region to obtain the parameters of the landmarks is omitted. Thus the direct regression provides a generic tool for plane parameter estimation. 

The MTL approach has proven to be beneficial in 2 out of the 4 body regions. For the case that a larger amount of data is available, we see further potential to reduce the error for all network architectures. Then, no substantial differences between the analyzed architecture variants are to be expected. However, the MTL approach will help to reduce the number of stored parameters and to facilitate a common network for standard plane regression. Thus, the network parameters need not to be loaded depending on the scanned body part, which saves time during the execution.
\\
\\
\noindent\textbf{Ethical Standards}: The data was obtained retrospectively from anonymized databases and not generated intentionally for the study. For this type of study formal consent is not required.

\noindent\textbf{Informed Consent}: The acquisition of data from living patients had a medical indication and informed consent was not required. The acquired data sets of cadavers were available retrospectively after they had been generated during surgical courses for physicians. The corresponding consent for body donation for these purposes has been obtained.

\noindent\textbf{Acknowledgements}: The authors gratefully acknowledge funding of the Erlangen Graduate School in Advanced Optical Technologies (SAOT) by the Bavarian State Ministry for Science and Art. This work was partially funded by Siemens Healthcare GmbH, Erlangen, Germany.

\noindent\textbf{Disclosures}: All authors declare that they have no conflict of interest.

\noindent\textbf{Disclaimer}: The methods and information presented here are based on research and are not commercially available.

\bibliographystyle{spiebib} 
\bibliography{3dAssessment} {}

\end{document}